\documentclass[aps,prb,twocolumn,groupedaddress]{revtex4-1}
\pdfoutput=1

\usepackage{graphicx}
\usepackage{amsmath,amssymb}

\usepackage{color}

\DeclareMathOperator{\acosh}{acosh}

\advance \textheight by 5\bigskipamount

\begin{document}

\title{Metamaterials with conformational nonlinearity}

\author{Mikhail Lapine}
\email[ Corresponding author: ]{ mlapine@uos.de }
\affiliation{Nonlinear Physics Centre, Research School of Physics and Engineering, Australian National University, Canberra ACT 0200, Australia}

\author{Ilya Shadrivov}
\affiliation{Nonlinear Physics Centre, Research School of Physics and Engineering, Australian National University, Canberra ACT 0200, Australia}

\author{David Powell}
\affiliation{Nonlinear Physics Centre, Research School of Physics and Engineering, Australian National University, Canberra ACT 0200, Australia}

\author{Yuri Kivshar}
\affiliation{Nonlinear Physics Centre, Research School of Physics and Engineering, Australian National University, Canberra ACT 0200, Australia}



\begin{abstract}
Within a decade of fruitful developments, metamaterials became a prominent
area of research, bridging theoretical and applied electrodynamics,
electrical engineering and material science.
Being man-made structures, metamaterials offer a particularly useful playground
to develop novel interdisciplinary concepts.
Here we demonstrate a novel principle in metamaterial assembly which
integrates electromagnetic, mechanical, and thermal responses within their elements.
Through these mechanisms, the conformation of the meta-molecules changes, providing a dual mechanism
for nonlinearity and offering nonlinear chirality.
Our proposal opens a wide road towards further developments of nonlinear
metamaterials and photonic structures, adding extra flexibility
to their design and control.
\end{abstract}

\maketitle


Metamaterials --- artificial materials designed to deliver an unusual electromagnetic
response --- have already established a prominent area of theoretical and experimental
physics\cite{SolSha,MarMarSor}
with applications ranging from super-imaging\cite{Pen0} and transformation optics\cite{PenSchSmi06}
to tunable and active materials\cite{BoaGriKiv11}, photonics\cite{Sha07} and plasmonics\cite{Bro10}.
At the same time, the corresponding ideas were readily adopted in acoustics,
promising unusual mechanical behaviour\cite{LiCha4,Nor08}. Clearly, simultaneous
access to electromagnetic, mechanical and thermal properties offers a great
application capabilities, as demonstrated, e.g. by mechanically tunable\cite{LapPowGor09}
and thermally reconfigurable\cite{TaoStrFan09} metamaterials.
Such a connection has been so far achieved by engineering the metamaterial
structure so as to provide an artificial means to control
electromagnetic properties; this approach is useful for tunability,
however, it does not provide a dynamic interaction.

\begin{figure}
\centering
\includegraphics[width=0.99\columnwidth]{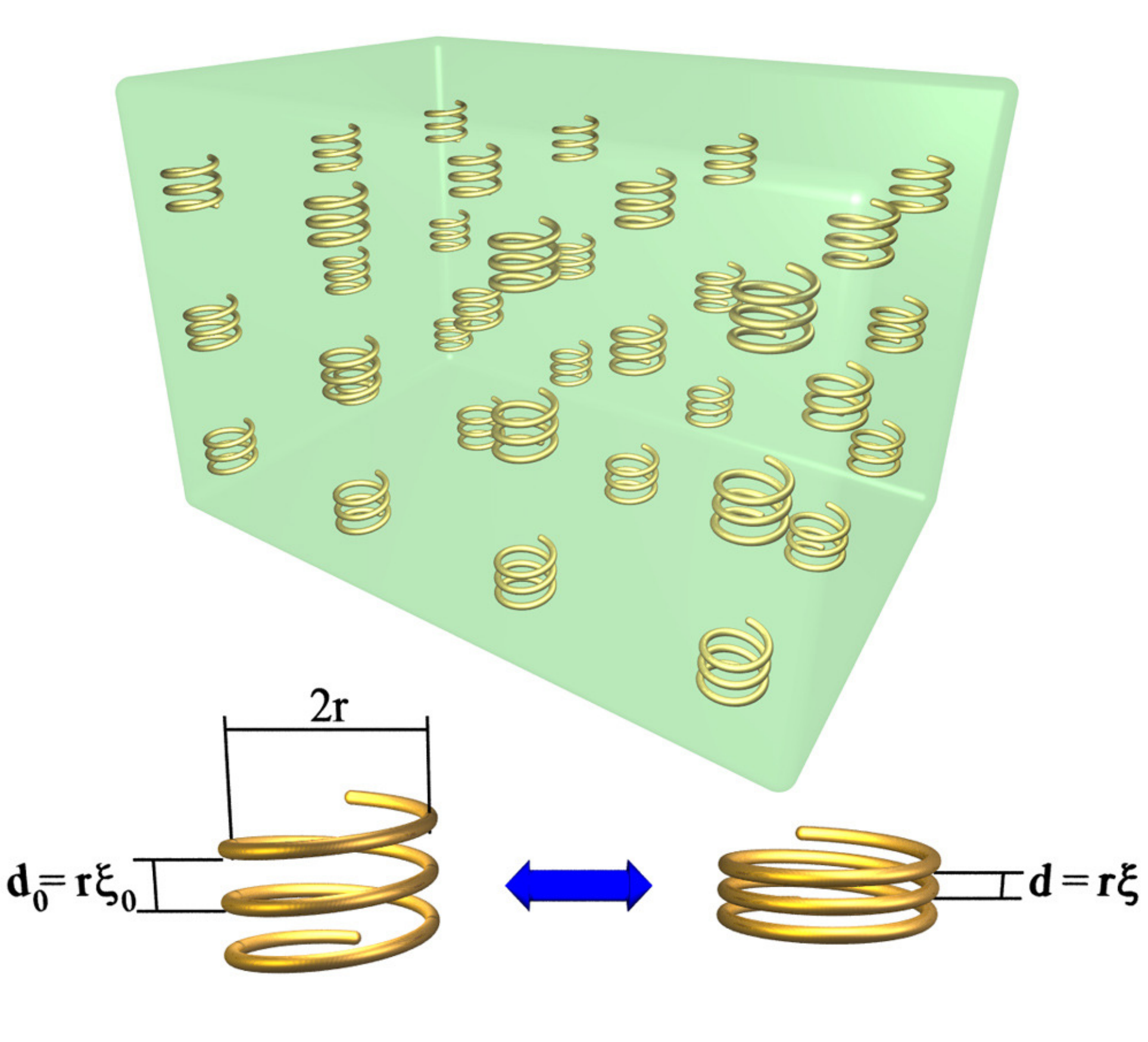}
\caption{\label{F1}
Schematic of a metamaterial composed of spiral meta-molecules, which are electromagnetic
resonators which can change their geometrical conformation: spiral pitch $\xi$ can vary, following
a balance between the attractive force induced by magnetic field and the spring rigidity;
and spiral radius $r$ can change upon thermal expansion.}
\end{figure}

Here we design a simple metamaterial element where responses of a different nature
are intrinsically coupled through its very structure.
Such an element is readily found as a thin wire wound into a spiral (Fig.~\ref{F1}).
From an electromagnetic perspective, it is a resonator with chiral properties,
and from a mechanical perspective it is a spring.
Finally, its specific shape makes all the characteristics sensitively depending on temperature.
This natural ``coincidence'' is of a great use as soon as the parameters of the
spiral allow thermo-magneto-mechanical coupling, which results
in efficient changes of its conformation.
Indeed, an external electromagnetic
wave induces a current along the spiral conductor, but this current causes an attractive
force between the windings. The spiral will therefore contract until that
force is balanced by the spring force, however the corresponding change in spiral geometry
will shift the resonance and thus alter the current amplitude in a self-consistent manner.
In addition, the thermal expansion of the spiral, resulting from its heating with
the increasing incident power, provides a relevant contribution, further shifting the
resonance --- importantly, in the same direction.
This dual conformational feedback leads to exotic nonlinear behaviour.

Spirals are chiral resonant particles known for about a century\cite{Lin14},
being suitable for producing bi-isotropic\cite{LSTV} or bi-anisotropic\cite{SSTS} media,
depending on how they are arranged in space.
Approximate models were developed to derive polarizability of simplified
canonical spirals\cite{TMS96}, which can be used to derive effective parameters
of dilute media, as well as for bi-anisotropic lattices of infinite spirals\cite{BST3}.
However, in the general case, the analysis of a periodic or random lattice of volumetric
spiral scatterers eludes analytical treatment. For this reason, in the qualitative analysis
in this paper we assume that the individual elements are arranged in such a way as
to minimize mutual interaction so that the mutual
coupling would only manifest itself in a small shift of the resonance frequency\cite{GLS2}.
We can therefore retrieve all the qualitative features by analysing
the response of an individual spiral.

%

Nonlinear behaviour of the metamaterials made of spiral meta-molecules
can be theoretically understood as its resonance frequency
essentially depends on the spiral pitch $\xi$
as well as on the radius of the spiral $r$.
Accordingly, the resonance shifts when the spiral compresses in response
to attraction between the windings as well as when a thermal expansion occurs
due to heating, both the effects triggered by the induced currents.
This ensures the intensity-dependent phenomena, providing for essential nonlinearity,
for example, in dependence of the induced magnetization on the incident magnetic field.
We note that the change in spiral pitch $\xi$ manifests itself with the corresponding change of
its chiral properties.
We can qualitatively describe the degree of chirality with the ratio between the electric $p_z$
and magnetic $m_z$ moments, induced in each spiral along its axis $z$.
This ratio is directly determined by the geometry of the spiral~\cite{BST3},
$ |\gamma| = \left| p_z / m_z \right| = {\xi} / {\omega \pi r} $.

Precise electromagnetic modelling of a multi-turn finite spiral is known to be
analytically challenging, however for a good qualitative description of the
nonlinear phenomena, we can rely on a simple circuit model \cite{BMM4}
which is applicable to two-windings spiral resonators if their
resonance frequency is well within the quasi-static regime (see \emph{Methods}).

For small $\xi$, the Amp{\`e}re force acting between the windings of the spiral,
can be calculated as that between the two parallel wires of the corresponding length.
Given the actual current distribution along the entire spiral \cite{BMM4},
we have found the net total force acting between the windings to be
$ F_\text{u} = \bigl. {\mu_0 I^2} \bigr/ {12 \xi} $.
This attractive force is be countered by the spring force $F_\text{s}$
(see \emph{Methods}).
The mechanical equilibrium is thus governed by
\begin{equation}
3 {G r^2 w^4} (\xi - \xi_0) \xi + 2{\mu_0 I^2} = 0,
\label{feq}
\end{equation}
which may seem to be a quadratic equation for $\xi$, however it is in fact far more
complicated as the current $I$ also depends on $\xi$ and $r$ through the impedance equation:
\begin{equation}
\left( R(r) + \mathrm i \omega L(r) - \frac{\mathrm i}{\omega C(r,\xi)} \right) \cdot I
= - \mathrm i \omega \mu_0 \pi r^2 H_0.
\label{imp}
\end{equation}
Here, $H_0$ is the amplitude of the magnetic field of the incident wave
(we imply the incident polarization with $\mathbf{H_0}$ being parallel to the spiral axis).

The thermal behaviour of the spiral is provided by its thermal expansion $r(T)$
as well as by the temperature dependence of its conductivity (see \emph{Methods}).
The amount of heat $Q^{+} = R I^2 / 2$ which is realized per second in the spiral,
is balanced, in a thermal equilibrium, to the heat loss through its surface
$Q^{-} = \beta \cdot 2 w (2 \pi r)^2 \cdot \Delta T$, where $\beta$ is a an empirical
coefficient which depends on the cooling conditions.
Thus, equations \eqref{feq}, \eqref{imp} and
\begin{equation}
\beta \cdot 2 w (2 \pi r)^2 \cdot \Delta T = R I^2 / 2
\label{tepl}
\end{equation}
form a system of coupled equations,
which can be numerically solved for $\xi$, $\Delta T$ and $I$ at a given frequency $\omega$
and amplitude $H_0$ of the incident field.


For a physical representation of the arising nonlinearity,
we can evaluate the total magnetic moment, induced in each spiral, and
obtain the effective magnetization of the metamaterial, $M = I \nu \pi r_0^2$
(where $\nu$ is the volumetric concentration of spirals).

\begin{figure}[t]
\centering
\includegraphics[width=0.99\columnwidth]{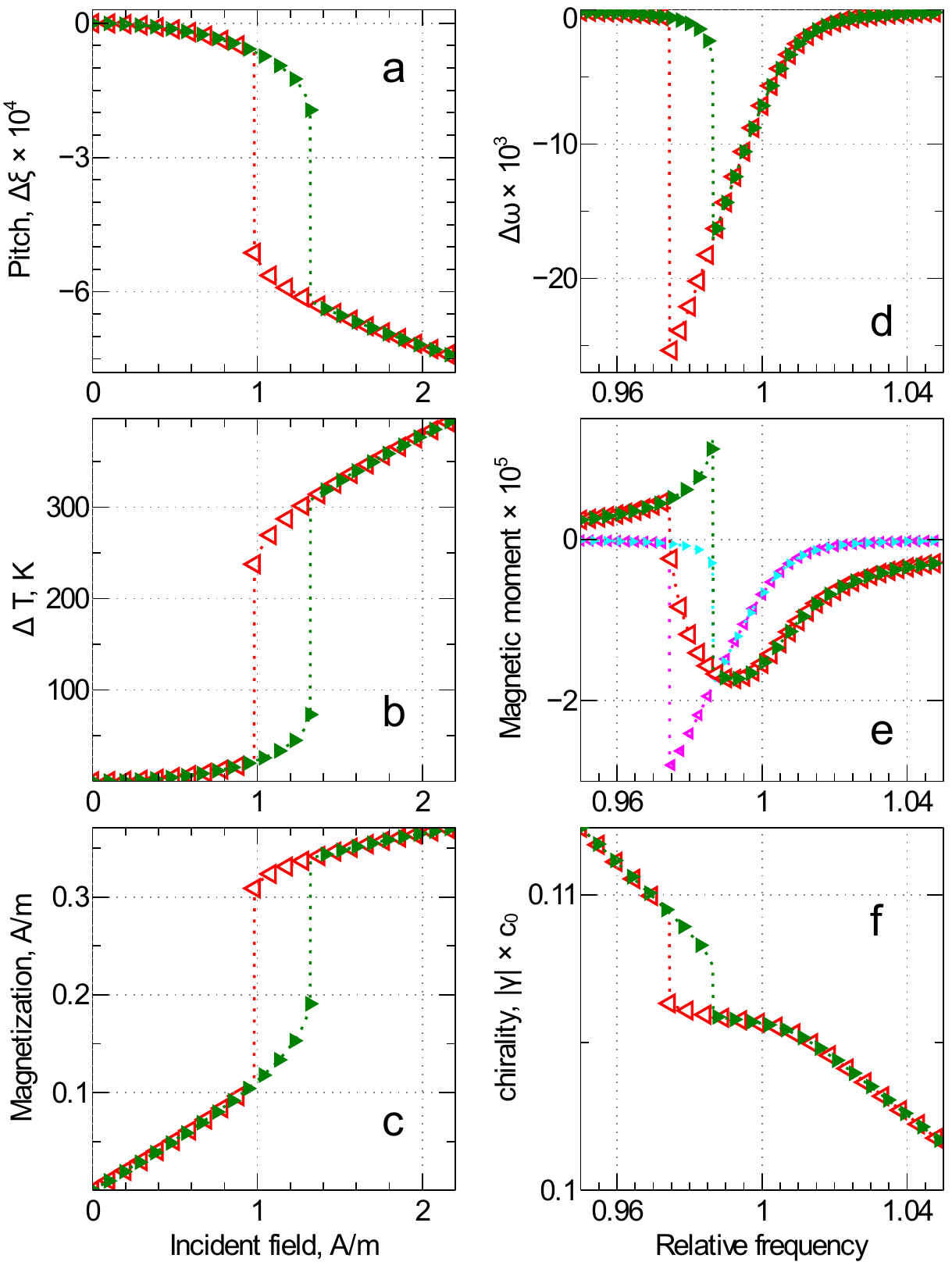}
\caption{\label{F2}
Dependence of the metamaterial properties on either (a-c) the incident amplitude $H_0$ (at the relative
frequency of $0.99\omega_0$), or (d-f) on the signal frequency (with the amplitude 2\,A/m),
for relaxed (right-pointing green triangles) and compressed (left-pointing red triangles) conformations.
Panel (a) shows the change in spiral pitch $\Delta \xi$;
panel (b) the increase in spiral temperature $\Delta T$,
panel (c) the absolute value of the effective magnetization,
panel (d) the relative shift in the resonance frequency $\Delta \omega$,
panel (e) the magnetic moment (real and imaginary parts) of a single spiral,
and panel (f) the chirality parameter $\gamma \cdot c_0$.
}
\end{figure}

In Fig.~\ref{F2}\,a--c, we illustrate the amplitude-dependent phenomena observed
when the signal frequency is slightly lower than $\omega_0$.
With the increase of power, the spiral compression enhances
the resonance, and therefore a positive feedback occurs upon a certain threshold, when
the spring force becomes insufficient to hold the resonantly growing attraction
[follow the green right-pointing triangles in Fig.~\ref{F2}a].
The spiral then jumps to another conformation, where a stronger compression
shifts the resonance sufficiently below the signal frequency, so that the induced force
becomes weak again, and further increase in power causes slow growth in magnetization.
However, if the field intensity is decreased from the high values, the spiral remains
stable in a compressed conformation much longer, as it is now trapped near the resonance.
It is only released when the peak amplitude of force is insufficient to hold it
[follow the red left-pointing triangles in Fig.~\ref{F2}a].
Further contribution is provided by the thermal expansion of the spiral
heated by the currents (Fig.~\ref{F2}b).
The two effects act simultaneously, providing a remarkable hysteresis-like pattern
for all the macroscopic properties of the metamaterial, such as magnetization
(Fig.~\ref{F2}c).

Two conformations can be also observed when the frequency of the incident radiation
is varied (Fig.~\ref{F2}\,d--f).
When the signal frequency increases at a given amplitude, the spiral remains just
slightly compressed until, close to the initial resonance $\omega_0$,
the attractive force becomes strong enough to trigger a jump to a compressed conformation
(so that the spiral resonance goes below the signal frequency), and further frequency
increase slowly relaxes this compression.
With decreasing frequency, the resonance of the spiral is already below the signal,
so approaching it steadily increases attractive forces and the spiral undergoes
continuous compression, led by the slope of the resonance.
To illustrate the frequency-dependent phenomena, we show how the resonance frequency
(Fig.~\ref{F2}d), the magnetic moment (Fig.~\ref{F2}e),
and the chirality (Fig.~\ref{F2}f) vary in hysteresis-like loops.


\begin{figure}[t]
\centering
\includegraphics[width=\columnwidth]{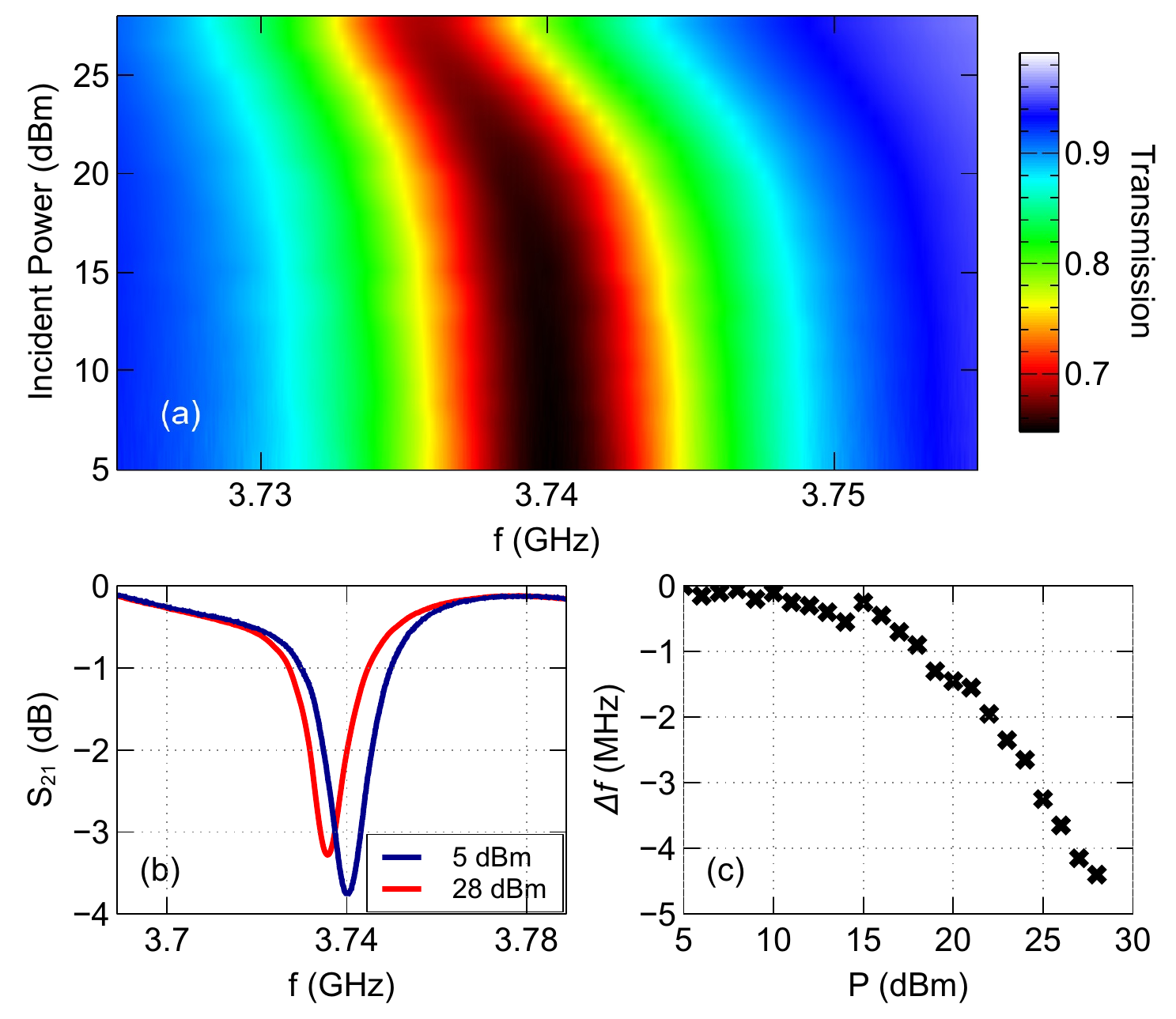}
\caption{\label{fig:experimental} Experimental results.
Panel (a) shows transmission through the spiral as a function of frequency and input power,
panel (b) the curves for the highest and lowest measured power,
panel (c) the frequency shift as a function of input power.}
\end{figure}

For the proof-of-principle experiments in the microwave range,
we take a thicker wire, such that the thermal effect dominates.
Fig.~\ref{fig:experimental}(a) shows the transmission amplitude as a function of input frequency and power. It can
be seen that there is a substantial shift of the resonance to lower frequencies as the input power is increased.
As can be seen in Fig.~\ref{fig:experimental}(b), the maximum frequency shift is less than the width of the resonances,
which is insufficiently strong to observe bistability. Figure~\ref{fig:experimental}(c)
shows a shift of the resonant frequency as a function of the input power.
Analysis of this shift allows to estimate the coefficient in Eq.~\eqref{tepl} which determines
the effective heat loss from the spiral, which turns out to be about $\beta \approx 30$.
We have used this coefficient for the above theoretical calculations.


The above results demonstrate that the flexible spiral resonator
exhibits nonlinear behaviour with two stable conformations of its meta-molecules.
Clearly, to ensure sufficiently strong force between the current loops, the
windings of the spiral have to be close to each other, $\xi \ll 1$, while
the wire should be thin enough to make the spring soft to respond to the
relatively weak Amp{\`e}re forces.

Among the parameters of the spiral, the relative wire thickness $w$ has the most drastic effect on the
interplay between the mechanical response (which increases as the fourth power of $1/w$)
and the thermal effect (largely independent of $w$).
We conclude that for the spiral's compression to dominate over its thermal expansion
the relative wire width should be smaller than $0.01$.
The conformational dynamics of the spiral is most interesting when the thermal and mechanical reactions
are comparable to each other, as shown with our theoretical examples.

Our proposal provides a starting platform for further analysis of the
nonlinear phenomena in various metamaterials based on the flexible chiral particle,
which can be made bi-isotropic, bi-anisotropic and even non-chiral, by choosing
an appropriate lattice and alignment.
While a lot of peculiar features will be provided with the arising mutual interactions,
the essence of nonlinear response is determined by the properties of an
individual spiral particle, and therefore the above analysis forms a reliable
basis for further research as well as the development of useful applications.


\footnotesize

\subsection*{METHODS}

In our theoretical analysis, $r$ is the radius of the spiral loops,
$r_w$ is the radius of the wire and $d$ is the
vertical distance between windings; the dimensionless normalized quantities
$w = r_w / r$ and $\xi = d / r$ are used for convenience (Fig.~\ref{F1}).

\begin{figure}[b]
\centering
\includegraphics[width=0.75\columnwidth]{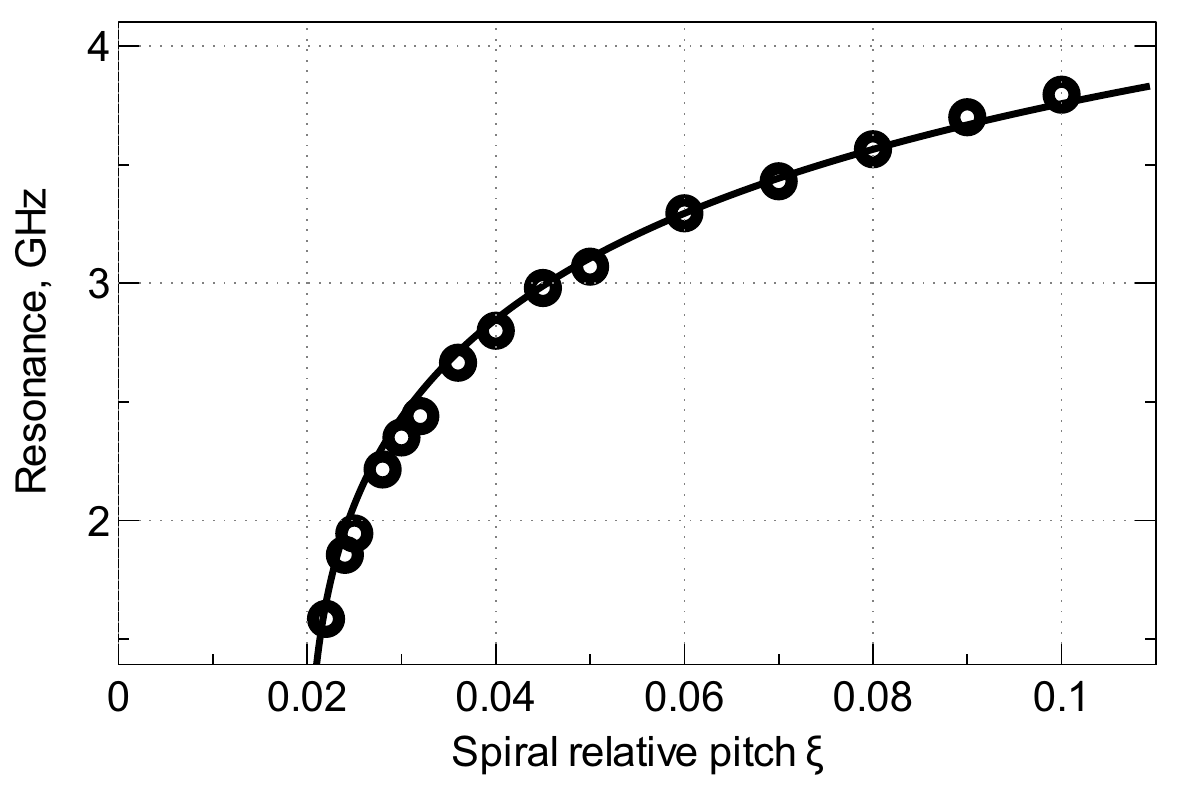}
\caption{\label{fig:spiral_freq} Resonance frequency of spiral resonators with different
pitch $\xi$. Comparison between the effective circuit theory (solid line)
and numerical simulations (circles). Spiral parameters here are $r_0 = 2$\,mm and $w = 0.01$.}
\end{figure}

With the quasi-static model\cite{BMM4} the spiral is described as an $LC$-circuit,
with the inductance $L$ being the same as the inductance of a single circular loop of
the same wire, $L = \mu_0 r \bigl( \ln (8/w) - 2 \bigr)$
(because the net current, flowing within the entire spiral ---
the sum over the current distribution in all turns ---
is uniform along the entire circumference).
The latter equally applies to the resistance $R = \bigl. \sqrt{\mu_0 \omega / 2 \sigma} \bigr/ w$;
both the $L$ and $R$ are taken with skin-effect in mind.
The capacitance $C$ is defined by that between the two windings
(with a minor correction for their curvature being negligible for small $\xi$):
$C(\xi) = \bigl. {2 \pi \varepsilon_0 \cdot \pi r} \bigr/ {\acosh (\xi / 2 w ) }$.
We have checked, with the aid of numerical simulations, that this simple circuit model
correctly predicts the resonant frequency $\omega_r (\xi) = 1/\sqrt{L(r) C(r,\,\xi)}$
of two-turn spirals (Fig.~\ref{fig:spiral_freq}).

The mechanical properties of a two-windings spiral are described by the stiffness coefficient
$ k = \bigl. {G r w^4} \bigr/ {8}$,
where $G$ is the shear modulus of the spiral material.
Note that the characteristic frequency of mechanical oscillations,
$\omega_M = \bigl. w \sqrt{3 G / 2 \rho} \bigr/ (4 \pi r)$ is many orders of magnitude
smaller than the electromagnetic frequencies involved,
so the analysis of the spiral geometrical reconfiguration is essentially static
and is determined by time-averaged amplitudes of the current.
The spring response is then described with the Hooke's law,
so it linearly increases with the deviation from the initial pitch value $\xi_0$:
$F_\text{s} = k r (\xi - \xi_0)$.

Temperature dependence of the spiral parameters is determined by a thermal
expansion coefficient $\alpha$ as $r = r_0 (1 + \alpha \cdot \Delta T)$.
Note that it leaves $\xi$ and $w$ unchanged
provided that the spiral is perfectly annealed and is free from internal stress.
The electric conductivity is affected in a similar way,
$\sigma = \sigma_0 (1 + \eta \cdot \Delta T)$
with a temperature coefficient of conductivity $\eta$.

For theoretical calculations, we assumed $r_0=5$\,mm,
$\xi_0=0.04$, $w=0.01$, and
the material parameters of copper for
$\sigma=6.6\cdot10^{7}$~S,
$G=40$\,GPa,
$\alpha = 1.7 \cdot 10^{-5}~\text{K}^{-1}$ and
$\eta = 4.3 \cdot 10^{-3}~\text{K}^{-1}$.

Numerical simulations were performed using the finite-element based
frequency-domain solver of CST Microwave Studio.
The spiral was excited by a plane wave, with its magnetic field parallel
to the axis of the the spiral.
The material was modelled as a perfect electrical conductor,
and a curved mesh was used to ensure good resolution of the geometry
without excessively fine meshing. The spiral was excited over a broad
frequency range, and the resonant frequency was determined as that with
the maximum magnitude of the normal magnetic field at the centre of the spiral.

For our experiments, the spirals were made of 100\,$\mu$m-thick copper wire
with 2.4\,mm radius and 0.5\,mm pitch.
We suspend a spiral parallel to the axis of a WR~229 rectangular waveguide
on a dielectric rod so that the incident magnetic field is along the spiral axis.
The waveguide is excited by a vector network analyzer (Rhode and Schwartz ZVB-20)
after amplification by a 1\,W amplifier (HP~83020A).
The output of the waveguide is measured by the network analyzer,
through a 20\,dB attenuator to eliminate gain compression in the detectors.
To ensure that the nonlinear response of the amplifier does not contribute
to the measurements, a power calibration is performed for each individual power level,
thus ensuring that the power level at the waveguide input is compensated for the
frequency and power dependent gain of the amplifier.
A directional coupler is used to sample the output from the amplifier,
so that the power calibration can be performed \textit{in situ},
and to provide a reference level against which to compare the output signal from the waveguide.
The input power to the waveguide was varied between 5\,dBm and 28\,dBm in steps of 1\,dB.


\subsection*{Acknowledgements}

The authors are grateful to Constantin Simovski, Stanislav Maslovski,
and Pavel Belov for useful discussions.
This work was supported by the Australian Research Council.



%

\end{document}